\documentclass[graybox]{svmult}

\usepackage{graphicx}
\usepackage{multicol}  
\usepackage[bottom]{footmisc}  
\usepackage{url}  

\usepackage{type1cm}  
\usepackage{newtxtext}
\usepackage[varvw]{newtxmath}  
\usepackage{makeidx}   
\makeindex
\usepackage{marvosym}  

\usepackage[numbers,sort&compress]{natbib}

\begin{document}

\title*{NASA's Cold Atom Laboratory: Five Years of Quantum Science in Space}
\titlerunning{NASA's CAL: Five Years of Quantum Science in Space}  

\author{
Kamal Oudrhiri 
\and James M. Kohel  
\and Nate Harvey \and James R. Kellogg \and David C. Aveline \and Roy L. Butler 
\and Javier Bosch-Lluis  
\and John L. Callas \and Leo Y. Cheng \and Arvid P. Croonquist \and Walker L. Dula 
\and Ethan R. Elliott  
\and Jose E. Fernandez \and Jorge Gonzales \and Raymond J. Higuera \and Shahram Javidnia \and Sandy M. Kwan \and Norman E. Lay \and Dennis K. Lee \and Irena Li \and Gregory J. Miles \and 
Michael T. Pauken \and Kelly L. Perry \and Leah E. Phillips \and Sarah K. Rees 
\and Matteo S. Sbroscia  
\and Christian Schneider  
\and Robert F. Shotwell \and Gregory Y. Shin \and Cao V. Tran \and Michel E. William 
\and Jason R. Williams  
\and Oscar Yang \and Nan~Yu 
\and Robert J~Thompson  
\and Diane C. Malarik \and DeVon W. Griffin \and Bradley M. Carpenter \and Michael P. Robinson 
\and Kirt Costello
}
\authorrunning{K. Oudrhiri, J. M. Kohel \textit{et al.}}  

\institute{
Kamal Oudrhiri (\Letter) \and James M. Kohel \and Nate Harvey \and James R. Kellogg \and David C. Aveline \and Roy L. Butler \and Javier Bosch-Lluis \and 
John L. Callas \and Leo Y. Cheng \and Arvid P. Croonquist \and Walker L. Dula \and Ethan R. Elliott \and Jose E. Fernandez \and Jorge Gonzales \and 
Raymond J. Higuera \and Shahram Javidnia \and Sandy M. Kwan \and Norman E. Lay \and Dennis K. Lee \and Irena Li \and Gregory J. Miles \and 
Michael T. Pauken \and Kelly L. Perry \and Leah E. Phillips \and Sarah K. Rees \and Matteo S. Sbroscia \and Christian Schneider \and 
Robert F. Shotwell \and Gregory Y. Shin \and Cao V. Tran \and Michel E. William \and Jason R. Williams \and Oscar Yang \and Nan Yu \and 
Robert J. Thompson 
\at Jet Propulsion Laboratory, California Institute of Technology, Pasadena, California \\ \email{kamal.oudrhiri@jpl.nasa.gov} 
\and 
Diane C. Malarik \and DeVon W. Griffin \and Bradley M. Carpenter \and Michael P. Robinson 
\at NASA Science Mission Directorate, Biological and Physical Sciences Division 
\and 
Kirt Costello \at NASA Johnson Space Center, Houston, Texas
}
\maketitle

\abstract{
NASA's Cold Atom Laboratory (CAL) is a multi-user science facility for studying quantum gases in the microgravity environment of the International Space Station. 
The persistent microgravity environment of the ISS enables research with ultra\-cold atoms in a temperature regime and force-free environment inaccessible to terrestrial laboratories,
unlocking the potential to observe novel quantum phenomena. 
CAL launched to the ISS in May 2018, 
and has operated continuously since then as the world's first multi-user quantum science facility in space. 
CAL is the first experimental facility to produce the fifth state of matter known as a Bose-Einstein condensate with ultra\-cold rubidium atoms on orbit \cite{Aveline2020} 
and, more recently, with mixtures of rubidium and potassium \cite{Elliott2023}.
We present an overview of CAL's design and operation, review the scientific contributions to date, and 
discuss recent on-orbit upgrades to extend its useful mission lifetime and provide enhanced science. 
We also consider opportunities for follow-on missions informed by lessons learned from over five years of operation on orbit.
}

\section{Introduction}

\begin{figure}[t]
	\centering\includegraphics[width=0.96\linewidth]{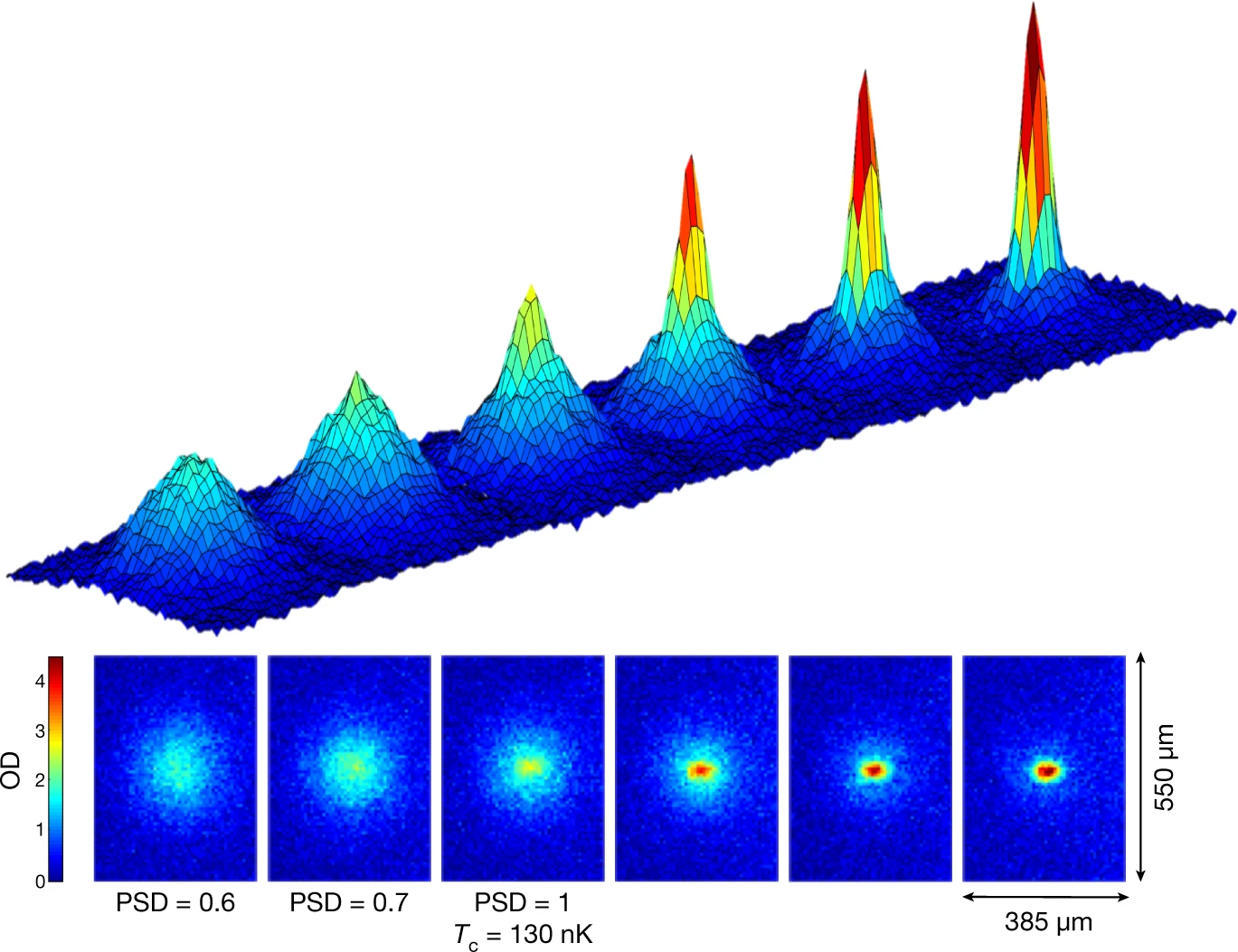}  
	\caption{Onset of Bose--Einstein condensation of rubidium atoms on the ISS \cite{Aveline2020}. 
	Each false-color image represents a separate experiment where atoms are released after evaporative cooling in a harmonic trap, 
	then imaged following a short time of free expansion to reveal the velocity distribution of the atomic ensemble.
	The final image shows a macroscopic cloud of almost 50\,000 atoms with over one quarter in a single quantum wave function determined by 
	the initial conditions in the trap.}
	\label{fig:RbBEC}
\end{figure}

Cold atom experiments in space are poised to revolutionize our understanding of physics in the coming decades. 
Among the myriad of proposed experiments are probes of the nature of the quantum vacuum, tests of quantum theories of gravity, investigations of novel quantum matter, and searches for dark energy and dark matter \cite{QST2023}. 
Space-based cold atom technologies offer the possibility for creating quantum sensors of unprecedented sensitivity, 
and practical applications abound, ranging from using atom interferometry to monitor the effects of climate change to developing space-based optical clocks that can synchronize timekeeping worldwide. 

The Cold Atom Laboratory (CAL) is the first experimental facility for the study of unique quantum-engineered states of matter 
in the micro\-gravity environment of the International Space Station \cite{Aveline2020}. 
This multi-user facility is the culmination of over three decades of rapid scientific and engineering development 
which has enabled the deployment of laboratory-based techniques to generate ultra\-cold atomic gases into space \cite{Aveline2020,Becker2018,Lachmann2021}.
CAL has reported the first on-orbit production of the quantum state of matter known as a Bose-Einstein condensate (BEC) with rubidium~\cite{Aveline2020} (Fig.~\ref{fig:RbBEC})
and also with dual-species mixtures of rubidium and potassium~\cite{Elliott2023}. 
A BEC is formed, in simplest terms, when atoms with integer spin are cooled below a critical temperature where the individual atoms' de~Broglie wavelengths become comparable to their mean separation; 
at this point the indistinguishable particles begin to condense into a single macro\-scopic wave\-function corresponding to the lowest accessible quantum state. 
The condensed atoms exhibit collective behavior in response to perturbations, allowing researchers to investigate quantum effects on a macroscopic scale using precisely controllable interactions with light, magnetic or electro\-magnetic fields~\cite{Ornes2017}. 
Since installation on the ISS in June 2018, CAL has operated for over five years on orbit, performing more than 100\,000 such experiments with ultra\-cold atoms
while traveling over 800 million miles.  

\begin{figure}[t]
  \centering
  \textbf{a}\hspace{-2em}\includegraphics[width=0.48\linewidth]{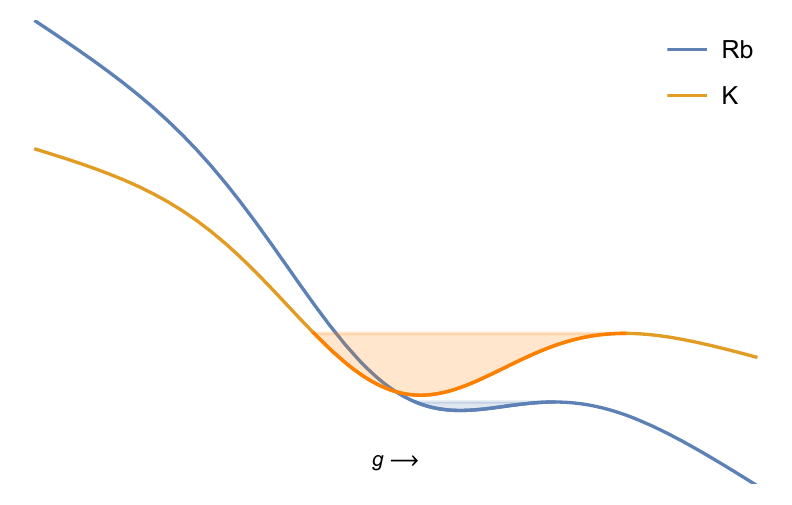} \hfill
  \textbf{b}\hspace{-2em}\includegraphics[width=0.48\linewidth]{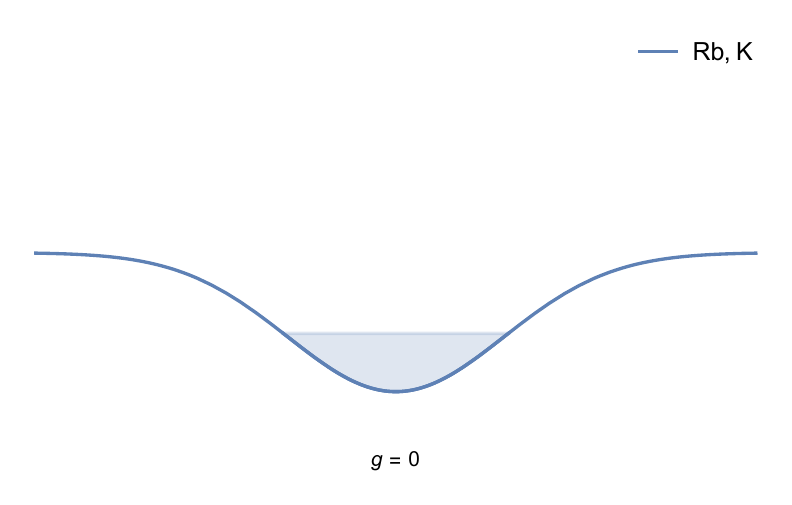}
  \caption{Ultra\-cold atoms in harmonic trap potentials aligned with gravity (\textbf{a}) on Earth and (\textbf{b}) in micro\-gravity.
  The micro\-gravity environment allows the use of much shallower potentials 
  and optimal overlap in mixtures of different atomic species.}
  \label{fig:MicrogravityBECs}
\end{figure}

The persistent free-fall environment on the ISS provides several compelling advantages for the production and study of ultra\-cold gas mixtures. 
The micro\-gravity environment 
allows the use of much weaker traps as compared to ground-based experiments (Fig.~\ref{fig:MicrogravityBECs}) and, as a result, the realization of even colder temperatures where quantum effects are magnified. 
Longer observation times are also possible, as the free expansion time within the measurement region isn't limited by gravitational acceleration.
For measurements of inertial forces, such as acceleration or gravity, the sensitivity scales as the \emph{square} of the observation time, giving space-based quantum sensors a dramatic advantage over their terrestrial counterparts. 
Micro\-gravity also enables optimal overlap of mixtures of different atomic masses without the need to compensate for the differential gravitational sag with applied magnetic field gradients.

Microgravity, however, is but one reason to deploy instruments with ultra\-cold atom in space. 
The cold temperatures and vacuum of deep space provide intriguing possibilities for pushing limits of ultra\-cold experiments well beyond anything that could be achieved on Earth~\cite{Thompson2023b}. 
Vast distance scales are accessible, as well, and experiments can be performed in a variety of reference frames and gravitational potentials.  
Finally, the solar system provides a wide variety of observational targets for quantum sensing instruments.

In the following, we present an overview of the CAL science mission after five years of operation on orbit. 
This chapter is derived from a paper presented at the SpaceOps 2023 Conference \cite{SpaceOps2023} 
and is organized as follows: 
Section \ref{Science} highlights CAL's scientific contributions to date, 
Section \ref{Design} describes the instrument design and operations, while 
Section \ref{ORUs} summarizes changes made on orbit since launch. 
Section \ref{Future} briefly discusses future plans for CAL and beyond.

\section{Science Achievements\label{Science}}

The Cold Atom Laboratory utilizes the microgravity environment of the ISS to study ultra\-cold quantum gases at unprecedented low energies and long free-fall times. 
CAL's science objectives were derived from the 2011 NASA Decadal Survey for Life and Physical Sciences \cite{Decadal2011}, and 
the facility offers investigators the ability to perform experiments with combinations of three different atomic species 
(${}^{87}$Rb with ${}^{41}$K or ${}^{39}$K) 
and to prepare them in specific atomic states (or superpositions of states).  
Atoms can be confined in a variety of trapping potential geometries, and dressed with both RF and microwave fields.  
Imaging of each species can be performed along two orthogonal directions, and 
interactions between atoms can be precisely tuned by varying an applied bias field over a magnetic Feshbach resonance~\cite{Chin2010}. 
Finally, light pulses from a far off-resonant laser at 785 nm (chosen so that it interacts with equal strength for both Rb and K atoms) can be applied for dual-species atom interferometry experiments. 
The instrument was designed not just to demonstrate these tools, which will be needed for a wide variety of future missions, but also to enable a variety of unique experiments that can only be performed in microgravity.

CAL was designed as a versatile multi-user science facility, enabling a world-class group of scientists to perform a diverse range of investigations of quantum phenomena in the microgravity environment of the ISS.
A NASA Research Announcement (NRA) was released in July 2013 to solicit proposals from academic and research institutions to utilize the Cold Atom Lab facility. 
From this NRA, 
five flight Principal Investigators (PIs) \cite{BigelowCALNRA,LundbladCALNRA,CornellCALNRA,SackettCALNRA,WilliamsCALNRA} 
and two ground PIs \cite{StamperKurnCALNRA,RaithelCALNRA} 
were selected. 
Among the selected PI teams are three recent Nobel Prize laureates.

\begin{figure}
    \centering\includegraphics[width=\linewidth]{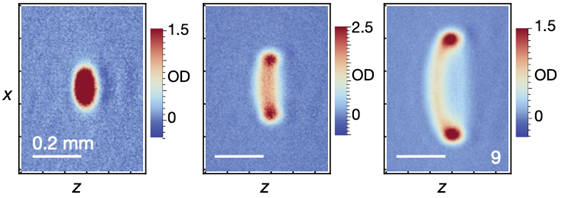}
    \caption{Absorption images of non-condensed rubidium atoms after release from a shell potential in micro\-gravity.
    The series of images illustrates the behavior as the bubble-shaped potential is ``inflated'' prior to release of the atoms. 
    The near-uniform densities are only observable in the absence of gravity.
    The darker lobes at the upper and lower bounds of each cloud are artifacts of the column-averaged absorption imaging technique
    combined with the finite imaging resolution.
    Adapted from Ref.~\citenum{Carollo2022}.}
    \label{fig:Bubbles}
\end{figure}

PI-led investigations have demonstrated the production of quantum ``bubbles'' in 
two-dimensional shell geometry traps \cite{Carollo2022,Lundblad2023}, 
as shown in Figure~\ref{fig:Bubbles}.
Other experiments have demonstrated adiabatic cooling in extremely weak traps in micro\-gravity~\cite{Pollard2020}, 
while others have employed advanced ``shortcut-to-adiabaticity'' protocols  
and delta-kick cooling techniques  
to achieve temperatures as low as 52~pK~\cite{Gaaloul2022}, 
corresponding to free-expansion velocities of about 100~\textmu{m}/s, 
with unprecedented precision in positioning cold atomic samples. 
Ongoing investigations include experiments to 
study the formation of Efimov molecules in micro\-gravity, 
demonstrate unique methods to correlate the positions of atoms, 
demonstrate a quantum rotation sensor, and 
search for novel phenomena involving mixtures of quantum gases.

\begin{figure}
  \centering\includegraphics[width=0.8\linewidth]{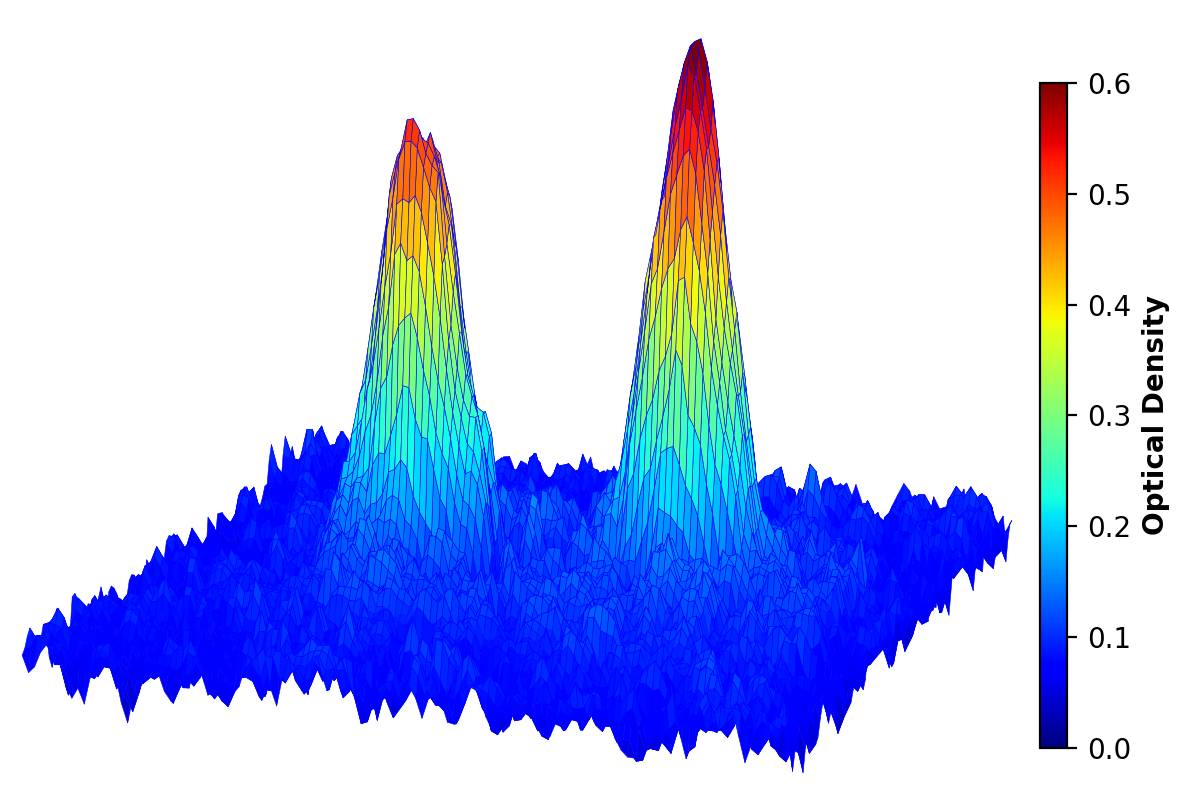}
  \caption{Superposition of momentum states observed in ultra\-cold rubidium atoms after applying a series of optical pulses to realize an atom interferometer. 
  Each spatially-separated atom cloud is approximately 40~{\textmu}m by 48~{\textmu}m in size, and 
  the clouds are separated in momentum space by two photon recoils.
  Prior to the observation, an individual atom's wave\-function exists simultaneously in both locations.}
  \label{fig:RbBraggAI}
\end{figure}

Atom interferometry is a particularly important application for cold atoms in space, and is an essential component in three of the five CAL PI science campaigns. 
In an atom interferometer, cold atoms serve as matter waves, while a laser light field creates the periodic grating structures that the atoms scatter from in order to realize a closed-loop \emph{matter wave} interferometer. 
This is in contrast to a traditional light interferometer, where photons behave as waves that diffract off of a physical structure.  
Figure~\ref{fig:RbBraggAI} shows a typical image resulting from the atom interferometer in CAL, illustrating the macroscopic separation of the two quantum superposition states for each atom. 
Beyond a simple demonstration of atom interferometry, CAL PIs have applied this quantum interference measurement technique for a proof-of-principle photon recoil measurement 
and to observe the influence of matter-wave interference over hundreds of milliseconds in free\-fall \cite{Williams2024}. 
CAL has also recently demonstrated simultaneous atom interferometers in dual-species mixtures of ${}^{87}$Rb and ${}^{41}$K~\cite{Elliott2023}.  
Efforts to increase the interaction time (and sensitivity) in this dual-species interferometer are ongoing, 
with the goal of employing differential interferometry for a proof-of-principle test of Einstein's equivalence principle. 

\section{Instrument Design and Operations\label{Design}}

\subsection{ISS Accommodation\label{ISS}}

As a space-based platform, the ISS offers a relatively benign environment for scientific payloads. 
The Cold Atom Laboratory was designed to fit into an EXPRESS (EXpedite the PRocessing of Experiments to Space Station) Rack 
that provides standardized power, mechanical, thermal, and data interfaces for scientific payloads on the ISS\@. 
CAL occupies one full ``quad'' locker plus a single locker within EXPRESS Rack 7 (ER-7) inside the U.S.\ Destiny Module, located near the station's center of gravity.
The instrument draws up to 565~W of power from the station's 28~VDC power 
with thermal management provided by water and forced-air cooling. 
A communications port supports daily real\-time science operations and continuous telemetry monitoring on the ground by the CAL Operations Team.

The CAL Science Instrument, which includes the Science Module as well as the majority of the lasers, optics and control electronics to support the proposed research programs, 
is housed within the ER-7 quad locker. 
These hardware subsystems are described in Section~\ref{FHW}.

The DC power conversion electronics are housed separately in a single locker in ER-7, along with an additional laser to support dual-species atom interferometry and the optical amplifier used for laser cooling potassium.

\subsection{Operational Concept\label{ConOps}}

\begin{figure}[t]
    \centering\includegraphics[width=\linewidth]{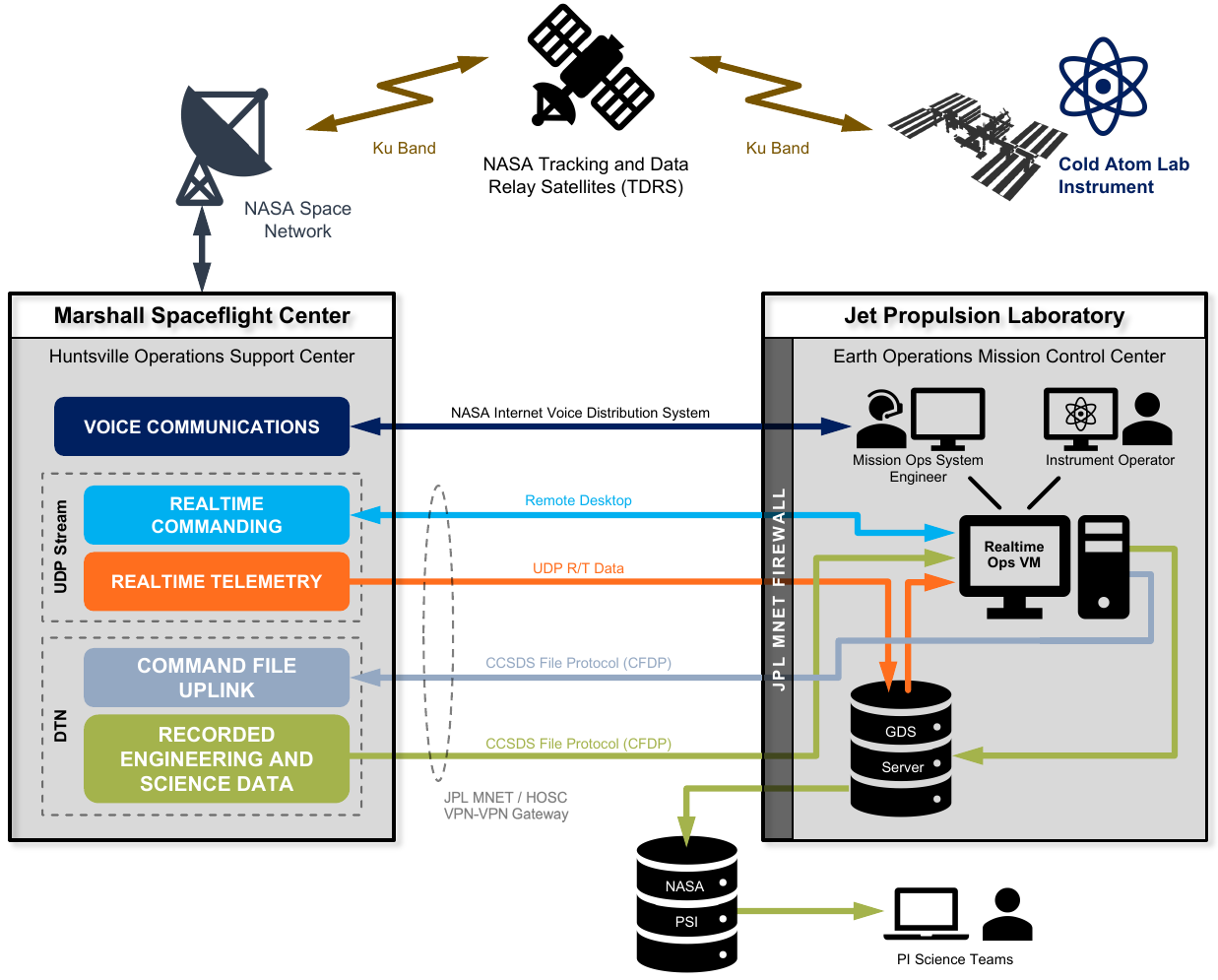}
    \caption{CAL Mission Operations Architecture.}
    \label{fig:MOSGDS}
\end{figure}

The CAL Operations Team operates the CAL Flight Instrument from the Earth Orbiting Missions Operations Center at JPL\@. 
Communication with the payload is over the Ku-band IP service through ISS Payload Operations at the Huntsville Operations Support Center (HOSC) at Marshall Space Flight Center (MSFC). 
The ground-to-station data link is provided by the Tracking and Data Relay Satellite System (TDRSS), a network of communications satellites and ground stations used to provide a near-continuous real\-time communications relay with the ISS\@. 
Data to and from the instrument are routed via two pipelines: 
For recorded engineering and science data as well as command file uplinks, 
CAL uses the delay tolerant networking (DTN) functionality provided by MSFC's Tele\-science Resource Kit (TReK) software suite. 
TReK's CCSDS File Delivery Protocol (CFDP) utility provides a standardized transport mechanism for file transfers over the DTN\@. 
For real\-time commanding, 
CAL uses a UDP stream from the ISS provided by HOSC's Ku-Band UDP Direct Downlink service. 
This UDP stream also serves as the pipeline for the live health and status telemetry flow from the instrument.
These data pathways are illustrated in 
Fig.~\ref{fig:MOSGDS}.

The operator interface to the Flight Instrument is provided via a Remote Desktop session on the ground computer, 
and experimental definition tables are executed on the Flight Instrument via sequence control by the CAL flight software. 
Science definition tables are developed by the PI Science Teams working with the CAL Team at JPL, and all new tables and sequences are flight rule checked before upload to the Flight instrument. 
Once on the Flight Instrument, the instrument operator queues the science table into the flight software's ``Sequence Engine'' 
where it can then be executed according to a time series of commands as specified in the corresponding time sequence file. 

A typical experimental sequence for single-species (Rb-only) science on CAL proceeds as follows: 
\begin{enumerate}
\item \textbf{Laser cooling}: Collect and cool atoms in a magneto-optical trap (MOT) inside the science region of the vacuum enclosure, followed by a brief stage of so-called ``optical molasses'' where the quadrupole magnetic field is turned off and the laser frequencies further detuned to reach atom temperatures below 100~\textmu{K}. 
Typical atom numbers for Rb are $N \approx 3 \times 10^8$ after this stage.
\item \textbf{State preparation}: Optically pump the cooled atoms to the low magnetic field-seeking quantum state. 
\item \textbf{Transfer to atom chip}: An intermediate quadrupole magnetic trap is used to transfer atoms from the MOT region to the atom chip-based trap at the top of the science region.
\item \textbf{Evaporative cooling}: An RF or microwave field is employed to eject the hottest Rb atoms from the chip trap by selectively transferring these atoms from the low field-seeking state to a high field-seeking state. 
The RF or microwave frequency is reduced over approximately 1.5~s to eject atoms at decreasing temperatures in a process known as ``forced evaporative cooling.'' 
At a critical temperature 
$T_\mathrm{c} \approx 100$~nK, 
atoms begin to collectively occupy a macroscopic Bose-condensed phase. 
\item \textbf{Decompression and release}: The atom trap is relaxed to further cool the atoms, then atoms are released and allowed to freely expand.
\item \textbf{Interrogation}: Atoms may be further probed using precise laser, magnetic, or RF pulses, as specified in the science definition table.
\item \textbf{Detection}: After a specified time of flight, an image of the expanded atom cloud is recorded by a camera using laser absorption imaging, followed by a reference image recorded after the destructive absorption image.
\end{enumerate}
Additional details related to dual-species operation, including sympathetic cooling in Rb/K gas mixtures, can be found in Refs.~\citenum{Elliott2023} and \citenum{Elliott2018}.

The primary science product generated with each table execution is the pair of absorption and reference images recorded at a specified time of flight. 
These images are transferred automatically from the Flight computer via CFDP to the ground computer, 
where the operator can review both the raw absorption images and the calculated optical densities in real\-time using image analysis software developed at JPL for this purpose. 
All science data is subsequently uplinked to NASA's Physical Science Informatics\footnote{\url{https://www.nasa.gov/PSI}} (PSI) archive for delivery to the PI Science Teams.

\subsection{Flight Hardware Overview\label{FHW}}


The CAL Flight Instrument includes a custom dual-species science module along with the necessary lasers, optics and control electronics to generate and manipulate ultra\-cold atoms for the various PI Science campaigns. 
The individual hardware subsystems are described in the following sections.

\subsubsection{Science Module\label{SM}}

\begin{figure}[t]
   	\centering\includegraphics[width=\linewidth]{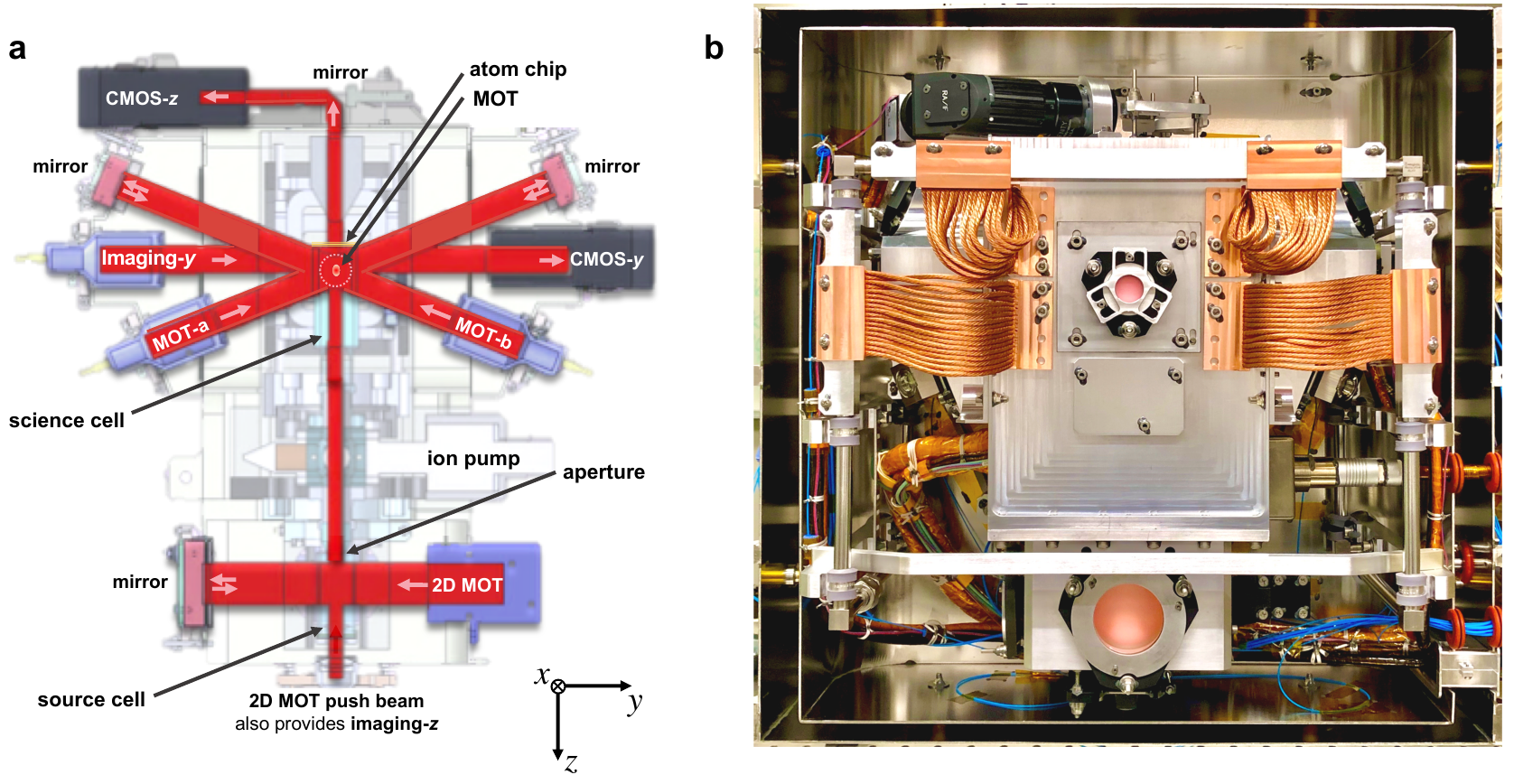}  
    \caption{(\textbf{a}) Illustration of the optical beam geometry in CAL's science module and 
    (\textbf{b}) an image of the science module during assembly, prior to installation of the front magnetic shields.
    Originally published in Ref.~\citenum{Aveline2020}.}
    \label{fig:SM}
\end{figure}

The CAL Science Module (Fig.~\ref{fig:SM}) includes a physics package containing Rb and K atoms in an ultra\-high vacuum (UHV) enclosure, 
along with an opto-mechanical bench that supports the surrounding laser beam collimators and free-space optics, cameras, RF and microwave antennas, and current coils for magnetic field control. 
The cameras and current coils are coupled to a water-cooling loop via flexible copper heat pipes for thermal control. 
A dual-layer magnetic shield encloses the entire science module and provides greater than 55~dB attenuation of external magnetic fields. 

The physics package is derived from ColdQuanta's commercial RuBECi chamber \cite{ColdQuanta}, 
modified for dual-species (Rb and K) operation and ruggedized for flight. 
The CAL physics package also incorporates a custom silicon chip-based atom trap containing a high-quality through-chip window. 
The custom atom chip provides a unique configuration of conductive current traces for creating the magnetic trapping potential near the chip surface, as well as features for improved electrical, thermal, and mechanical integrity for use in flight.

The vacuum enclosure of the physics package consists of two distinct regions, both made with high optical quality glass walls, 
separated by a differential pumping aperture. 
The source region contains two \textit{in vacuo} alkali metal dispensers for Rb and K, while the UHV science region includes 
a miniaturized $2 \ell/\mathrm{s}$ ion pump plus a graphite non-evaporable getter to maintain background pressures below $10^{-10}$~Torr 
within this region to allow long trap lifetimes.
A two-dimensional MOT, created by two pairs of circularly-polarized laser beams along the orthogonal horizontal axes plus a two-dimensional quadrupole magnetic field, 
acts as an ``atom funnel'' to collect and transfer the slowest atoms from a dilute thermal vapor of Rb and K in the source region to the UHV science region through the aperture.

This collimated beam of laser-cooled Rb and K atoms is captured in the science region by a three-dimensional MOT formed by circularly-polarized laser beams along three orthogonal axes and centered on a quadrupole magnetic field formed by a pair of current coils in the anti-Helmholtz configuration. 
After further laser cooling followed by optical pumping to a pure magnetic field-sensitive state, the atoms are transferred to the atom chip-based magnetic potential, 
where RF or microwave forced evaporative cooling is employed to reach the transition to a BEC\@.
After release from the magnetic trap, ultra\-cold atoms are imaged using absorption imaging, where a laser pulse resonant with the atomic transition probes the density distribution of the expanded atom cloud to reveal the initial momentum state of the atomic ensemble.

Two imaging subsystems are provided within the Science Module to allow absorption imaging along orthogonal axes, 
either parallel or perpendicular to the atom chip. 
The wide-field imaging subsystem employs a large diameter (12 mm FW($1/e^2$)M) laser beam aligned just below the atom chip to provide a wide field of view.
The orthogonal through-chip imaging axis has a much smaller beam, and makes use of the high optical quality window at the center of the atom chip. 
Both cameras 
employ a near-infrared enhanced scientific CMOS sensor, 
with a quantum efficiency of approximately 35\% at the resonant absorption wavelengths for Rb and K\@. 

\subsubsection{Lasers and Optics Subsystem\label{LOS}}

The Lasers and Optics Subsystem in CAL performs the initial laser cooling and trapping, optical pumping, and resonant detection of Rb and K 
atoms within the Science Module. 
To accomplish this, two laser frequencies are required for each atomic species. 
These frequencies are generated by tunable narrow line\-width ``trapping'' and ``repumping'' lasers which are frequency-offset locked to a reference laser, which is in turn frequency-stabilized to a narrow atomic transition in a spectroscopy module containing a vapor cell of either rubidium or potassium.
This offset-lock scheme provides the required frequency agility from the trapping and repumping lasers for the laser cooling, optical pumping, and detection stages.
The tunable laser outputs are further amplified using two tapered-chip semiconductor amplifiers to provide up to 350~mW of optical power at the 780~nm and 767~nm wavelengths for Rb and K, respectively, 
and delivered to the Science Module through a polarizing-maintaining optical fiber-based network of optical switches and fiber splitters/combiners.
Beam delivery via fiber optics allows the placement of lasers and optical components outside the 
Science Module, and facilitates replacement of individual subassemblies during integration or, if necessary, on orbit. 

A separate fixed-wavelength laser at 785~nm generates the far-off-resonant light for dual-species atom interferometry using Bragg diffraction in an optical lattice. 
The interferometer pulse sequence is generated using an acousto-optic modulator (AOM) driven at the resonant radio\-frequencies for simultaneous Bragg diffraction of ${}^{87}$Rb and ${}^{41}$K (or ${}^{39}$K). 
The multiple frequencies for driving the AOM are directly synthesized by an arbitrary waveform generator, as described in the following section.

\subsubsection{Control and Electronics Subsystem\label{Electronics}}                                  

A Windows-based computer controller plus three field-programmable gate array (FPGA) modules, housed in a DC-powered PXI chassis, 
provide dynamic control of the magnetic field currents, RF and microwave emitter frequencies, and laser frequencies and amplitudes 
during each experimental sequence. 
The primary FPGA provides digital and analog timing waveforms for synchronous control of the current drivers, direct digital synthesizers, arbitrary waveform generator, and RF and optical switches with a timing resolution of 10~{\textmu}s.
The remaining two FPGAs are used to implement digital servo control loops to acquire and maintain the frequency-stabilization locks for the Rb and K reference lasers. 
Executive functions, such as loading and processing experimental configuration tables, running experimental control sequences, collecting and reporting real-time telemetry data, and monitoring for off-nominal conditions, are handled by the LabVIEW-based ``PXI Host'' flight software running as a Windows application on the PXI controller.
Two dozen lower-level hardware-control software modules directly interface with the various hardware subsystems and are managed by the PXI Host.

The Current Driver Assembly (CDA) contains independently-controllable low noise 
current drivers for the six magnetic field coils inside the science module 
and the three atom chip trap current traces, 
as well as additional current drivers for the Rb and K dispensers in the Science Module. 
Two of the three atom chip drivers are switchable across three different atom chip traces and can provide either unidirectional or bidirectional current (depending on the switch selection) to generate multiple magnetic trap configurations. 
The third current driver can be directed across a ``fast Feshbach'' current loop to generate a bias field up to 90~G in the Science Module.
This bias field is employed to access magnetic Feshbach resonances in mixtures of Rb and K atoms, where the sign and strength of interactions between atomic species can be precisely tuned by varying an external magnetic field.

The Laser Frequency Lock Assembly (LFLA) contains the laser control electronics for the six narrow-linewidth lasers 
as well as the ultra\-high frequency (UHF) and microwave frequency sources for driving atomic transitions in Rb or K\@.
The laser control electronics include all the laser drivers, frequency stabilization electronics and reference synthesizers 
for the offset-locked lasers, and are housed as six individually removable electronic ``slices'' within the LFLA chassis.

Frequency sources for evaporative cooling of atoms include an 80-MHz arbitrary waveform generator (AWG) 
housed within the PXI chassis, as well as three RF/UHF direct digital synthesizers (DDS) on LFLA Slices 7 and 8. 
Slice 7 also contains a phase-locked 7.3~GHz dielectric resonant oscillator (DRO) that is mixed with a 1~GHz DDS to generate microwave frequencies for evaporative cooling. 
LFLA Slice 9 contains the high-power RF amplifiers used with the AWG to drive either the RF loop antenna within the Science Module for evaporative cooling of Rb 
or the AOM to generate the optical Bragg diffraction pulses for atom interferometry.
An RF relay on Slice 9 directs the amplified signal to either subsystem.

\section{On-Orbit Upgrades\label{ORUs}}

CAL was designed to allow on-orbit replacement of limited-lifetime components, including lasers, optical amplifiers, and the alkali-metal dispensers inside the Science Module, in order to extend science operations beyond its three-year primary mission. 
To support the anticipated replacement of the identified hardware, the CAL payload launched in 2018 with a suite of on-orbit replacement unit (ORU) lasers and amplifiers, as well as additional modules for the PXI chassis. 
The CAL Operations Team continuously monitors telemetry from these lifetime-limited and consumable items to identify any change in performance indicating end-of-life behaviors, as well as to monitor the health and status of the various hardware subsystems. 

The ability of the ISS crew to access a scientific payload housed inside the station also enables the replacement, under certain conditions, of hardware due to unanticipated failures or degraded performance on orbit or, as in CAL's Science Module upgrade described in Section~\ref{sm3}, 
to deliver enhanced science capabilities to the instrument. 

\subsection{Enhanced Science Module\label{sm3}}

In December 2019, after 18 months of CAL's operation in orbit, a new atom-interferometry capable science module was delivered to the ISS on the SpaceX CRS-19 resupply mission. 
The upgraded science module, known as Science Module 3 (SM3), was designed and assembled at JPL to fully support planned 
experiments with atom interferometry in space, including a proof-of-principle test of Einstein's Equivalence Principle. 

During transport on ground and after unloading on the station, 
the ion pump in the science module was operated using a GSE ion pump controller assembly (IPCA), developed at JPL, 
to maintain vacuum integrity within the module's physics package. 
The science module can be stored greater than three months without power to its ion pump under normal conditions on ground, but 
this powered stowage was a precaution against vacuum degradation due to helium permeation into the glass-walled physics package in an elevated helium environment.

\begin{figure}[t]
	\includegraphics[width=0.5\linewidth]{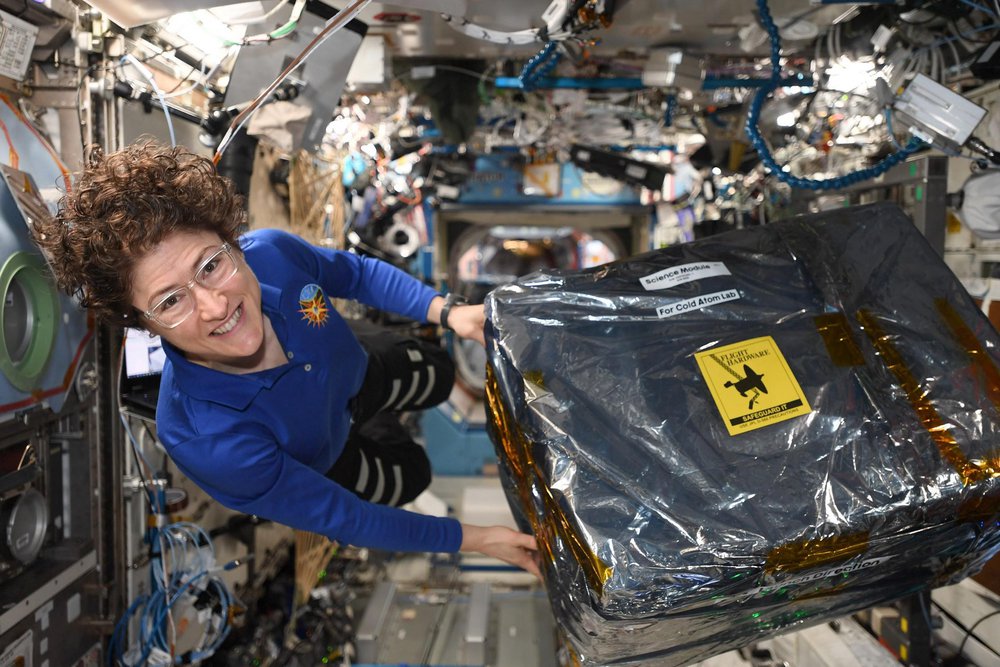}
	\includegraphics[width=0.5\linewidth]{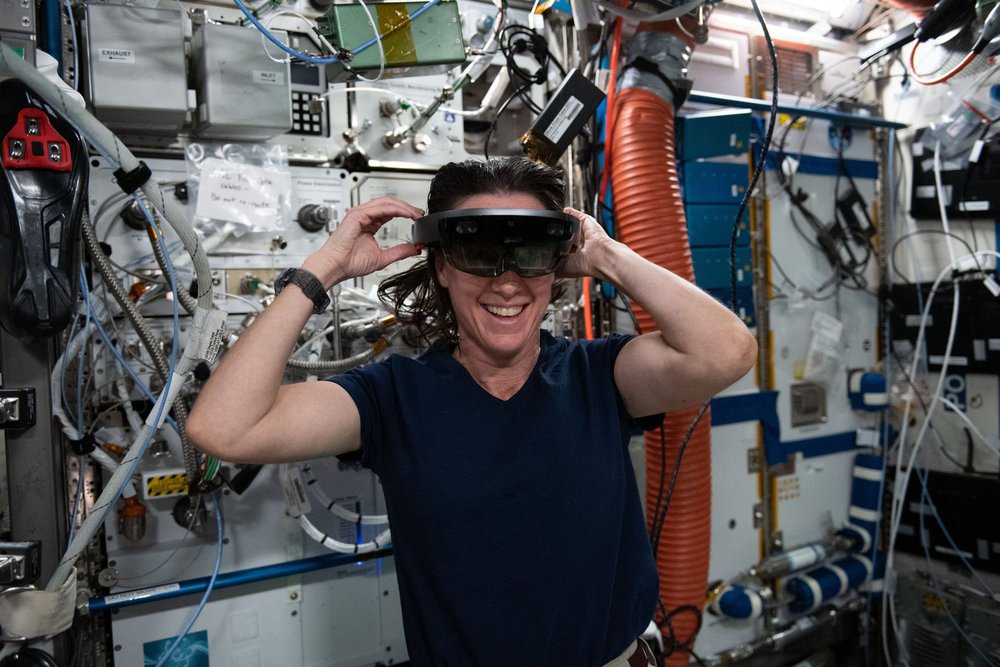}
	\caption{\emph{Left}: Astronaut Christina Koch unloads a new Science Module aboard the International Space Station in December 2019, 
	prior to installation in the Cold Atom Laboratory in January 2020.
	\emph{Right}: Astronaut Megan McArthur is shown wearing the augmented-reality headset used during installation of hardware inside the 
	Cold Atom Laboratory in July 2021.
	Images courtesy of NASA.}
	\label{fig:RnRs}
\end{figure}

The removal and replacement (R\&R) of CAL's science module was performed in January 2020 by the ISS crew (see Fig.~\ref{fig:RnRs}) under the direction of the CAL Operations Team.
This R\&R activity involved twelve separate crew procedures that took place on five days over a nine-day span. 
After closeout of the newly-installed SM3, the original science module, SM2, was connected to the GSE IPCA to maintain vacuum until its return to ground for analysis at JPL\@.
During SM2's return flight, the IPCA was continuously powered by a specialized lithium ion battery that had been flight-qualified for use on the station for extravehicular activities (EVAs) by the astronauts.

Immediately following this R\&R activity, the CAL Operations Team confirmed the vacuum integrity within the UHV science region of SM3 from instrument telemetry, then the Team proceeded to demonstrate laser-cooled atoms. 
After uploading new experimental definition tables developed 
on ground for the specific atom chip geometry in SM3, the CAL Team was able to confirm nominal and repeatable generation of BECs in the upgraded instrument.

\subsection{Upgraded Microwave Frequency Source\label{slice7b}}

In July 2021, the ISS crew upgraded the Cold Atom Laboratory with new hardware to enable the production of ultra\-cold potassium atoms 
alongside rubidium, as required for dual-species science operations. 
This hardware, referred to as ``Slice 7B'' and installed in the LFLA chassis, completed the microwave frequency synthesis chain required to directly cool rubidium atoms 
using evaporative cooling with microwave frequencies, rather than RF, 
which then sympathetically cool either ${}^{39}$K or ${}^{41}$K atoms within the same magnetic trap.
Previous experiments with rubidium relied solely on RF for evaporative cooling, which is far less efficient in dual-species mixtures with potassium \cite{Elliott2018}.

During the installation of this hardware, astronaut Megan McArthur employed a Microsoft HoloLens mixed and augmented reality headset (Fig.~\ref{fig:RnRs}), 
in a first demonstration of this technology to assist a crew procedure aboard the ISS\@. 
Preparation for this activity took six months and involved a collaboration between engineers at NASA's JPL, Johnson Space Center, and Marshall Space Flight Center.
Through a live video feed from the headset camera, the CAL Operations Team at JPL was able to share the astronaut's view of the hardware being replaced on orbit, 
and to simultaneously affix virtual text and graphical annotations alongside physical objects within the augmented reality environment to assist the installation in real time.
For example, during the detailed procedure of reconnecting each cable assembly as the new hardware was being installed, 
the CAL Team was able to place a cursor to indicate a specific connector or cable tie to supplement the written procedure. 
The virtual cursor would remain fixed relative to the indicated object, independent of the motion of the headset-mounted camera.

Following installation, the CAL Operations Team was able to validate the performance of this hardware by demonstrating evaporative cooling of rubidium atoms to a BEC using only microwave frequencies from Slice 7B\@. 
Subsequently, the CAL Team was further able to generate a Bose-condensed sample of ${}^{41}$K using only sympathetic cooling of potassium atom by microwave evaporatively-cooled rubidium atoms within the same trap.

\subsection{CPU Controller and SSD Replacement}

During science operations on 12 August 2021, the CAL Operations Team lost communication with the Flight Computer and were unable to reconnect. 
Efforts to ping the Flight Computer were unsuccessful, even after multiple remote power-cycles of the payload in an attempt to induce a reboot of the Flight Computer. 
From the available power draw telemetry, it was determined that the most likely causes were a failure of the PXI-8108 CPU controller or the solid-state drive (SSD) inside this controller, 
and/or a corruption of the Windows operating system on this drive.

An spare ORU controller had been in stowage on the station since the original payload delivery in 2018, 
and the ISS crew was able to remove and replace the original controller with this ORU on 28 August 2021. 
The newly-installed controller was then reconfigured by the ISS Network Team for operation on the ISS network.
After communication with ground was established, the latest flight software was remotely installed on the new controller, 
along with the TReK and ION DTN software suites, 
allowing CAL to resume operation on 3 September 2021. 

A subsequent R\&R procedure was necessary after the boot volume in this ORU controller became corrupted after less than three months of operation. 
The non-functional SSD was replaced on 16 December 2021 with a bootable drive that was delivered to the station on SpaceX CRS-23. 
Following the successful R\&R, the CAL Operations Team was again able to reconnect and remotely install the Flight Software on the newly installed drive. 
Once the FSW installation was verified, recent experimental control tables and sequences required to operate the Flight Instrument were re-uploaded. 
After confirming nominal telemetry from all hardware subsystems, the Operations Team proceeded with a successful checkout of the rubidium subsystem 
and was thereafter able to resume science operations.



\section{Future Plans\label{Future}}

As the CAL mission continues beyond its fifth year, additional hardware upgrades are planned to further enhance the science return and allow new categories of investigations.  
The ORU Science Module 3X is scheduled to launch in early 2026
This new science module should provide a significant increase in the number of ultra\-cold atoms, 
in part due to the inclusion of a dynamically-variable ``meso\-scale'' magnetic trap which allows improved transfer of laser-cooled atoms
from the initial trapping potential (formed by an external pair of coils in the Helmholtz configuration) 
to the orders-of-magnitude smaller atom chip-based micro\-trap.  
By dynamically varying the size and position of this intermediate trap to couple atoms between the initial and final magnetic traps, 
atom numbers on chip could improve by as much as an order of magnitude,
bringing CAL's performance in micro\-gravity on par with typical terrestrial experiments in terms of this metric. 

\subsection{Lessons Learned for Future Missions}\label{LessonsLearned}

Our experience over the past five years on orbit guides the development of science missions beyond CAL, and  
we can identify a number of design architecture and operational improvements which can improve both the utility and reliability of future space-based experiments with ultra\-cold atoms.

\begin{description}
\item[\textit{Faster experimental cycle times}:]
CAL typically operates on a 75--90~s experiment cycle, and is limited by thermal considerations to cycle times no shorter than 60~s.
Decreasing this cycle time to ten or even five seconds would enable a corresponding increase the science throughput of the instrument.

\item[\textit{Better experimental diagnostics}:]
Dynamic \textit{in situ} measurements of laser powers and both magnetic and RF field strengths would dramatically improve our ability to identify and diagnose systematic issues 
or degradation of hardware, and could aid scientific investigations. 

\item[\textit{More modular design}:]
The improved ability to quickly swap out hardware modules on orbit can vastly increase both the reliability and scientific versatility of an experimental facility on the station.  
A more modular architecture, combined with the increased availability of crew time that has come with the advent of a commercial crew, 
would facilitate on-orbit configuration changes 
and perhaps allow trained experimental physicists to conduct hands-on science in space.

\item[\textit{Operation from PI host institutions}:] Giving PI teams direct control of the instrument will allow them to develop and refine new experiments fluidly, similar to how atomic physics experiments are conducted on the ground. 
For this to become possible, it will be necessary to further automate the instrument to allow experiments to run autonomously 
and to incorporate the level of diagnostics necessary to ensure the safety of the instrument and the station crew.
\end{description}

\subsection{Follow-On Missions}\label{FO}

The Bose-Einstein Condensate Cold Atom Lab (BECCAL) \cite{Frye2021} is a complementary NASA-DLR quantum matter research facility expected to launch to the ISS after CAL completes its nominal mission in 2027. 
BECCAL is designed to provide a fast experimental duty cycle, unique dynamically-configurable trapping potentials, and novel capabilities for atom interferometry. 
In contrast to the CAL atom interferometer design, which uses a single far-detuned laser to interact simultaneously with two atomic species (Rb and K) for unprecedented common-mode suppression of vibration and noise in weak equivalence principle experiments, 
BECCAL will use two separate lasers for dual-species atom interferometry, 
with higher sensitivity to spurious vibrations but a factor of 10 larger atom beams and active control of the retro-optic 
to allow unprecedented long interrogation times. 
The advanced capabilities of BECCAL will enable new studies of non-linear atom optics, matter-wave cavities, gravity gradients, and tests of Newton's gravitational constant~\cite{Frye2021}. 

A proposed astronaut-operated Quantum Explorer \cite{Thompson2023a} follow-on to BECCAL would provide a reconfigurable facility for easy swap-out of custom hardware, PI-specific instruments, lasers, and science modules. 
Research enabled by such a facility could include the study of topics as diverse as the nature of the quantum vacuum; quantum chaos and pattern formation; atom lasers and matter-wave holography; matter-wave localization; and quantum simulations of astrophysical objects, such as the early universe, black holes, and neutron stars, as well as condensed matter systems such as high temperature superconductors.

\section{Conclusion\label{Conclusion}}

The Cold Atom Laboratory is a pathfinder mission for fundamental studies of quantum matter in micro\-gravity, 
and for future space-based quantum sensors enabling exquisitely precise measurements for both applied and fundamental science applications.
While there are micro\-gravity alternatives to an orbiting platform 
(e.g.\ parabolic flights \cite{Stern2009,Barrett2016}, 
drop towers \cite{vanZoest2010,Muntinga2013}, or 
sounding rockets \cite{Becker2018,Lachmann2021}), 
the science return can be much greater in an orbital facility 
where hundreds or even thousands of experimental sequences can be processed per day. 
As a multi-user facility, 
CAL was designed to support a diversity of experimental campaigns with multiple science teams, 
and to provide the flexibility to evolve as the supporting ground-based research matures. 
For an exploratory mission investigating a wide variety of quantum phenomena over an extended mission lifetime, these advantages were compelling.

CAL is also unique in that it is the only Flight mission simultaneously in Phases C/D (Design \& Development) through E (Operations), as the project  continues to develop and test hardware for on-orbit upgrades as it operates through an extended science-phase mission.
The ability to enhance science capabilities and replace limited-lifetime hardware is a unique advantage of the crewed ISS platform. 
CAL has undergone several on-orbit upgrades and repairs over its five years of operations with support from the station crew.  
These upgrades have enabled CAL to not only continue operations beyond its primary science mission 
but also provided \emph{enhanced capabilities for increased science return}. 
As part of one such upgrade, we have demonstrated the first use of 
augmented reality technology to improve the real-time interaction between an astronaut and payload engineers on ground during an instrument maintenance procedure.
This technology promises to find applications far beyond CAL for maintaining science payloads on the station.


\section*{Acknowledgments}
The Cold Atom Laboratory is supported by the Biological and Physical Sciences Division of NASA's Science Mission Directorate and the ISS Program Office. 
We thank Dr.~Craig Kundrot, former Division Director of NASA BPS, and Dr.~Ulf Israelsson, former Program Manager for the Fundamental Physics Office at JPL, for their long-term support.
We also acknowledge the early support of Dr.~Mark Lee, former Program Scientist for NASA BPS; 
without his vision the Cold Atom Laboratory would not have become a reality. 

CAL was designed, built, and is currently managed and operated by the Jet Propulsion Laboratory, California Institute of Technology, under a contract with the National Aeronautics and Space Administration. US Government sponsorship is acknowledged. 
\copyright~2024 All Rights Reserved.

\bibliographystyle{spbasic_unsrt}  
\bibliography{BEC, AtomInterferometry, Microgravity, QuantumGases}

\end{document}